\documentclass[12pt]{article}
\usepackage{epsfig}

\newcommand\pubnumber{DESY 01/003 \\   LU-ITP 01/01 \\
  PSI-PR-01-01\\UR-1630, ER/40685/965}
\newcommand\pubdate{\today}
\newcommand\hepnumber{hep-ph/0101257}

\def\support{\footnote{Heisenberg fellow of the Deutsche 
Forschungsgemeinschaft}} 

\def\Title#1{\begin{center} {\Large\bf #1 } \end{center}}
\def\Author#1{\begin{center}{ \sc #1} \end{center}}
\def\Address#1{\begin{center}{ \it #1} \end{center}}

\newcommand\pubblock{\rightline{\begin{tabular}{l} \pubnumber\\
         \pubdate\\ \hepnumber \end{tabular}}}
\newenvironment{Abstract}{\begin{quotation}  }{\end{quotation}}
\newenvironment{Presented}{\begin{quotation} \begin{center} 
             Presented at the\end{center}
      \begin{center}\begin{large}}{\end{large}\end{center} \end{quotation}}
\def\Acknowledgments{\bigskip  \bigskip \begin{center}
          \large\bf Acknowledgments\end{center}}

\makeatletter
\def\section{\@startsection{section}{0}{\z@}{5.5ex plus .5ex minus
 1.5ex}{2.3ex plus .2ex}{\large\bf}}
\def\subsection{\@startsection{subsection}{1}{\z@}{3.5ex plus .5ex minus
 1.5ex}{1.3ex plus .2ex}{\normalsize\bf}}
\def\subsubsection{\@startsection{subsubsection}{2}{\z@}{-3.5ex plus
-1ex minus  -.2ex}{2.3ex plus .2ex}{\normalsize\sl}}

\renewcommand{\@makecaption}[2]{%
   \vskip 10pt
   \setbox\@tempboxa\hbox{\small #1: #2}
   \ifdim \wd\@tempboxa >\hsize     
       \small #1: #2\par          
     \else                        
       \hbox to\hsize{\hfil\box\@tempboxa\hfil}
   \fi}

 \def\citenum#1{{\def\@cite##1##2{##1}\cite{#1}}}
 
\newcount\@tempcntc
\def\@citex[#1]#2{\if@filesw\immediate\write\@auxout{\string\citation{#2}}\fi
  \@tempcnta\z@\@tempcntb\m@ne\def\@citea{}\@cite{\@for\@citeb:=#2\do
    {\@ifundefined
       {b@\@citeb}{\@citeo\@tempcntb\m@ne\@citea\def\@citea{,}{\bf ?}\@warning
       {Citation `\@citeb' on page \thepage \space undefined}}%
    {\setbox\z@\hbox{\global\@tempcntc0\csname b@\@citeb\endcsname\relax}%
     \ifnum\@tempcntc=\z@ \@citeo\@tempcntb\m@ne
       \@citea\def\@citea{,}\hbox{\csname b@\@citeb\endcsname}%
     \else
      \advance\@tempcntb\@ne
      \ifnum\@tempcntb=\@tempcntc
      \else\advance\@tempcntb\m@ne\@citeo
      \@tempcnta\@tempcntc\@tempcntb\@tempcntc\fi\fi}}\@citeo}{#1}}
\def\@citeo{\ifnum\@tempcnta>\@tempcntb\else\@citea\def\@citea{,}%
  \ifnum\@tempcnta=\@tempcntb\the\@tempcnta\else
  {\advance\@tempcnta\@ne\ifnum\@tempcnta=\@tempcntb \else\def\@citea{--}\fi
    \advance\@tempcnta\m@ne\the\@tempcnta\@citea\the\@tempcntb}\fi\fi}
\makeatother

\def\beq{\begin{equation}}
\def\eeq#1{\label{#1}\end{equation}}
\def\eeqn{\end{equation}}

\newenvironment{Eqnarray}%
   {\arraycolsep 0.14em\begin{eqnarray}}{\end{eqnarray}}
\def\beqa{\begin{Eqnarray}}
\def\eeqa#1{\label{#1}\end{Eqnarray}}
\def\eeqan{\end{Eqnarray}}

\let\bar=\overbar

\def\ie{{\it i.e.}}

\def\M{{\cal M}}

\def\Dslash{\not{\hbox{\kern-4pt $D$}}}
\def\dslash{\not{\hbox{\kern-2pt $\del$}}}

\def\CM{{\mbox{\scriptsize CM}}}

\def\GF{G_F}

\def\msb{{\bar{\ssstyle M \kern -1pt S}}}

\def\lsim{\mathrel{\raise.3ex\hbox{$<$\kern-.75em\lower1ex\hbox{$\sim$}}}}
\def\gsim{\mathrel{\raise.3ex\hbox{$>$\kern-.75em\lower1ex\hbox{$\sim$}}}}

\usepackage{axodraw}
\usepackage{epsfig}

\makeatletter

\expandafter\ifx\csname mathrm\endcsname\relax\def\mathrm#1{{\rm #1}}\fi

\newcounter{saveeqn}

\@addtoreset{equation}{section}

\newcount\@tempcntc
\def\@citex[#1]#2{\if@filesw\immediate\write\@auxout{\string\citation{#2}}\fi
  \@tempcnta\z@\@tempcntb\m@ne\def\@citea{}\@cite{\@for\@citeb:=#2\do
    {\@ifundefined
       {b@\@citeb}{\@citeo\@tempcntb\m@ne\@citea
        \def\@citea{,\penalty\@m\ }{\bf ?}\@warning
       {Citation `\@citeb' on page \thepage \space undefined}}%
    {\setbox\z@\hbox{\global\@tempcntc0\csname
b@\@citeb\endcsname\relax}%
     \ifnum\@tempcntc=\z@ \@citeo\@tempcntb\m@ne
       \@citea\def\@citea{,\penalty\@m}
       \hbox{\csname b@\@citeb\endcsname}%
     \else
      \advance\@tempcntb\@ne
      \ifnum\@tempcntb=\@tempcntc
      \else\advance\@tempcntb\m@ne\@citeo
      \@tempcnta\@tempcntc\@tempcntb\@tempcntc\fi\fi}}\@citeo}{#1}}

\def\@citeo{\ifnum\@tempcnta>\@tempcntb\else\@citea
  \def\@citea{,\penalty\@m}%
  \ifnum\@tempcnta=\@tempcntb\the\@tempcnta\else
   {\advance\@tempcnta\@ne\ifnum\@tempcnta=\@tempcntb \else
\def\@citea{--}\fi
    \advance\@tempcnta\m@ne\the\@tempcnta\@citea\the\@tempcntb}\fi\fi}

\def\nl{\nonumber\\}

\renewcommand{\lsim}
{\mathrel{\raisebox{-.3em}{$\stackrel{\displaystyle <}{\sim}$}}}
\renewcommand{\gsim}
{\mathrel{\raisebox{-.3em}{$\stackrel{\displaystyle >}{\sim}$}}}
\def\asymp#1%
{\mathrel{\raisebox{-.4em}{$\widetilde{\scriptstyle #1}$}}}

\def\Nequal#1%
{\mathrel{\raisebox{-.5em}{$\stackrel{=}{\scriptstyle\rm#1}$}}}
\newcommand{\dsl}[1]{\not \hspace{-0.7mm}#1}
\def\dsl{\mathpalette\make@slash}
\def\make@slash#1#2{\setbox\z@\hbox{$#1#2$}%
  \hbox to 0pt{\hss$#1/$\hss\kern-\wd0}\box0}

\def\beq{\begin{equation}}
\def\eeq{\end{equation}}
\def\beqar{\begin{eqnarray}}
\def\eeqar{\end{eqnarray}}
\def\barr#1{\begin{array}{#1}}
\def\earr{\end{array}}
\def\bfi{\begin{figure}}
\def\efi{\end{figure}}
\def\btab{\begin{table}}
\def\etab{\end{table}}
\def\bce{\begin{center}}
\def\ece{\end{center}}
\def\nn{\nonumber}
\def\disp{\displaystyle}
\def\text{\textstyle}

\def\al{\alpha}

\def\ga{\gamma}

\def\veps{\varepsilon}
\def\la{\lambda}

\def\si{\sigma}

\def\refeq#1{\mbox{(\ref{#1})}}

\def\reffi#1{\mbox{Figure~\ref{#1}}}
\def\reffis#1{\mbox{Figures~\ref{#1}}}

\def\citere#1{\mbox{Ref.~\cite{#1}}}
\def\citeres#1{\mbox{Refs.~\cite{#1}}}

\newcommand{\TeV}{\unskip\,\mathrm{TeV}}
\newcommand{\GeV}{\unskip\,\mathrm{GeV}}
\newcommand{\MeV}{\unskip\,\mathrm{MeV}}

\newcommand{\ri}{{\mathrm{i}}}
\newcommand{\rd}{{\mathrm{d}}}

\def\mathswitchr#1{\relax\ifmmode{\mathrm{#1}}\else$\mathrm{#1}$\fi}

\newcommand{\PW}{\mathswitchr W}
\newcommand{\Pw}{\mathswitchr w}
\newcommand{\PZ}{\mathswitchr Z}

\newcommand{\Pe}{\mathswitchr e}

\newcommand{\Pd}{\mathswitchr d}

\newcommand{\Pu}{\mathswitchr u}

\newcommand{\Pt}{\mathswitchr t}

\newcommand{\Pep}{\mathswitchr {e^+}}
\newcommand{\Pem}{\mathswitchr {e^-}}
\newcommand{\PWp}{\mathswitchr {W^+}}
\newcommand{\PWm}{\mathswitchr {W^-}}

\def\mathswitch#1{\relax\ifmmode#1\else$#1$\fi}

\newcommand{\MW}{\mathswitch {M_\PW}}

\newcommand{\MZ}{\mathswitch {M_\PZ}}

\newcommand{\Me}{\mathswitch {m_\Pe}}

\newcommand{\GW}{\Gamma_{\PW}}
\newcommand{\GZ}{\Gamma_{\PZ}}

\newcommand{\sw}{\mathswitch {s_\Pw}}

\renewcommand{\GF}{\mathswitch {G_\mu}}

\def\solid{\raise.9mm\hbox{\protect\rule{1.1cm}{.2mm}}}
\def\dash{\raise.9mm\hbox{\protect\rule{2mm}{.2mm}}\hspace*{1mm}}

\def\ie{i.e.\ }

\newcommand{\CCOT}{{\mathrm{CC03}}}
\newcommand{\IBA}{{\mathrm{IBA}}}
\newcommand{\Born}{{\mathrm{Born}}}

\newcommand{\Coul}{{\mathrm{Coul}}}

\renewcommand{\min}{{\mathrm{min}}}

\newcommand{\recomb}{{\mathrm{rec}}}

\def\Im{\mathop{\mathrm{Im}}\nolimits}

\newcommand{\eeWW}{{\Pe^+ \Pe^-\to \PW^+ \PW^-}}

\newcommand{\eeWWffff}{\Pep\Pem\to\PW\PW\to 4f}
\newcommand{\eeffff}{\Pep\Pem\to 4f}

\newcommand{\kon}{\hat{k}}

\def\draftdate{\relax}
\def\mda{\relax}
\def\mua{\relax}
\def\mla{\relax}
\def\draft{
\def\thtystars{******************************}
\def\sixtystars{\thtystars\thtystars}
\typeout{}
\typeout{\sixtystars**}
\typeout{* Draft mode!
         For final version remove \protect\draft\space in source file *}
\typeout{\sixtystars**}
\typeout{}
\def\draftdate{\today}
\def\mua{\marginpar[\boldmath\hfil$\uparrow$]%
                   {\boldmath$\uparrow$\hfil}%
                    \typeout{marginpar: $\uparrow$}\ignorespaces}
\def\mda{\marginpar[\boldmath\hfil$\downarrow$]%
                   {\boldmath$\downarrow$\hfil}%
                    \typeout{marginpar: $\downarrow$}\ignorespaces}
\def\mla{\marginpar[\boldmath\hfil$\rightarrow$]%
                   {\boldmath$\leftarrow $\hfil}%
                    \typeout{marginpar: $\leftrightarrow$}\ignorespaces}
\def\Mua{\marginpar[\boldmath\hfil$\Uparrow$]%
                   {\boldmath$\Uparrow$\hfil}%
                    \typeout{marginpar: $\uparrow$}\ignorespaces}
\def\Mda{\marginpar[\boldmath\hfil$\Downarrow$]%
                   {\boldmath$\Downarrow$\hfil}%
                    \typeout{marginpar: $\downarrow$}\ignorespaces}
\def\Mla{\marginpar[\boldmath\hfil$\Rightarrow$]%
                   {\boldmath$\Leftarrow $\hfil}%
                    \typeout{marginpar: $\leftrightarrow$}\ignorespaces}
\overfullrule 5pt
\oddsidemargin -15mm
\marginparwidth 29mm
}

\def\stars{\strut\leaders\hbox{*}\hfill\strut}
\def\starline{\hfil\strut\hfil\hbox to \textwidth {\stars}\hfil}

\begin{document}
\begin{titlepage}
\pubblock

\vfill
\def\thefootnote{\fnsymbol{footnote}}
    \Title{ Off-shell W-pair production --- \\[5pt]
universal versus non-universal corrections%
}
\vfill
\Author{A.\ Denner}
\vspace*{-2.0em}
\Address{Paul Scherrer Institut,
CH-5232 Villigen PSI, Switzerland}
\Author{S.\ Dittmaier\support}
\vspace*{-2.0em}
\Address{Deutsches Elektronen-Synchrotron DESY, D-22603 Hamburg, Germany}
\Author{M.\ Roth}
\vspace*{-2.0em}
\Address{Institut f\"ur Theoretische Physik, Universit\"at Leipzig,
D-04109 Leipzig, Germany}
\Author{D.\ Wackeroth}
\vspace*{-2.0em}
\Address{Department of Physics and Astronomy, University of Rochester,
\\ Rochester, NY 14627-0171, USA}
\vfill
\begin{Abstract}
Electroweak radiative corrections to $\Pep\Pem$ scattering processes
typically amount to ${\cal O}(10\%)$ at LEP energies.
Their logarithmic increase with energy renders them even more
important at future colliders.
Although the bulk of these corrections is due to
universal process-independent effects, the remaining non-universal
corrections are nevertheless phenomenologically important.
We describe the structure of the universal corrections to $\eeWWffff$
in detail and discuss the 
numerical size of universal and non-universal effects 
\looseness 1
using the Monte Carlo generator {\sc RacoonWW}%
\footnote{Program available from http://www.hep.psi.ch/racoonww/racoonww.html}.
\end{Abstract}
\vfill
\begin{Presented}
5th International Symposium on Radiative Corrections \\ 
(RADCOR--2000) \\[4pt]
Carmel CA, USA, 11--15 September, 2000
\end{Presented}
\vfill
\end{titlepage}
\def\thefootnote{\arabic{footnote}}
\setcounter{footnote}{0}

\section{Introduction}
\label{se:intro}

At present,
the investigation of W-pair production at LEP2 plays an important role
in the verification of the Electroweak Standard Model (SM).
Apart from the direct observation of the triple-gauge-boson
couplings in $\eeWW$, the increasing accuracy in the W-pair-production
cross-section and W-mass measurements
has put this process into the row of SM precision tests \cite{wwexp}.
The W-pair cross section is measured at the per-cent level, and the
W-boson-mass determination aims at a final accuracy of $30\MeV$.
Experiments at a future $\Pep\Pem$ linear collider (LC) with higher luminosity
and higher energy will even exceed this precision.

To account for the high experimental accuracy on the
theoretical side is a great challenge: the W bosons have to be treated
as resonances in the full four-fermion processes $\eeffff$, and radiative
corrections need to be included. While several lowest-order predictions 
are based on the full set of Feynman diagrams, only very few calculations 
include radiative corrections beyond the level of universal radiative
corrections (see \citeres{lep2repWcs,lep2mcws} and references therein). 
These universal corrections comprise the leading
process-independent effects; for $\eeWWffff$ these include universal
renormalization effects (running or effective couplings), the Coulomb
singularity at the W-pair-production
threshold, and initial-state radiation (ISR)
in leading-logarithmic approximation. The remaining corrections are
usually viewed as non-universal
and can only be included by an explicit diagrammatic calculation.
In this article we describe the
structure of the universal corrections in detail and discuss the size of
the non-universal corrections for LEP2 und LC energies.
This issue is not only theoretically interesting, it is also important
in practice, since many Monte Carlo generators for W-pair production
that are in use neglect non-universal electroweak corrections.

The size of non-universal corrections was already estimated by
inspecting the pair production of stable W~bosons quite some time ago
\cite{lep2repWcs,bo92,di97}. For LEP2 energies these effects reduce
the total W-pair cross section at the level of 1--2\%, but for
energies in the TeV range the impact grows to ${\cal O}(10\%)$.
For differential distributions the size of the non-universal corrections
is usually much larger. In the following we investigate the
corresponding corrections to off-shell W-pair production, $\eeWWffff$,
by inspecting total cross sections as well as angular and
invariant-mass distributions with the Monte Carlo generator {\sc RacoonWW} 
\cite{de00}.

\section{Radiative corrections to off-shell W-pair production --- state
of the art}

Fortunately, to 
match the experimental precision for W-pair production a full one-loop 
calculation for the four-fermion processes is not needed for most purposes, 
in particular for LEP2 physics. Instead it is sufficient to take into 
account only those radiative corrections that are enhanced by two resonant 
W bosons. For centre-of-mass (CM) energies $E_{\CM}$ not too close to the 
W-pair-production
threshold, say for $E_{\CM}\gsim 170\GeV$, the neglected 
${\cal O}(\alpha)$ corrections are of the order $(\al/\pi)(\GW/\MW)$, \ie
below 0.5\% even if possible enhancement factors are taken into account.
The theoretically clean way to carry out this approximation is the 
expansion about the two resonance poles, which is called {\it double-pole 
approximation} (DPA). A full description of this strategy and 
of different variants used in the literature (some of them involving 
further approximations) can be found in \citeres{de00,yfsww,Be98,ku99}.

\begin{sloppypar}
At present, two Monte Carlo programs include ${\cal O}(\alpha)$ corrections 
to $\eeWWffff$ in DPA and further 
numerically important higher-order effects: 
{\sc YFSWW3} \cite{yfsww} and {\sc RacoonWW} \cite{de00}. The salient 
features of the two approaches, which are conceptually very different,
as well as detailed comparisons of numerical results
are summarized in \citere{lep2mcws}. Further numerical results of the
two programs can be found in \citeres{de99a,ja00}.
Both programs have reached an
accuracy of roughly $\sim 0.5\%$ for CM energies between $170\GeV$ and
$500\GeV$. For higher energies also leading 
electroweak two-loop effects become important (see also below).
\end{sloppypar}

\begin{figure}
\setlength{\unitlength}{1cm}
\centerline{
\begin{picture}(10.5,10.5)
\put(-.5,-.8){\includegraphics{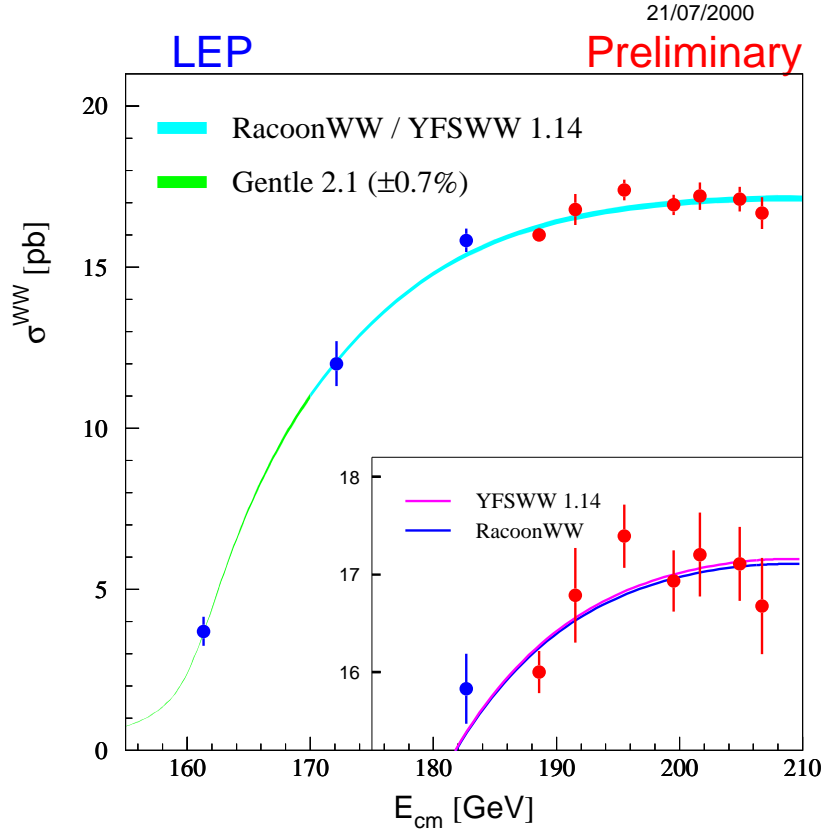}}
\end{picture}
}
\caption{Total WW production cross section at LEP2 as given by the
LEPEWWG \cite{LEPEWWG}}
\label{fig:wwcs2000}
\end{figure}

Figure~\ref{fig:wwcs2000} shows a comparison of the results of 
{\sc RacoonWW} and {\sc YFSWW3} with recent LEP2 data, as given 
by the LEP Electroweak Working Group \cite{LEPEWWG} for the Summer 
2000 conferences.
The data are in good agreement with the predictions of the two programs,
which differ by about 0.3\% at LEP2 energies.
Below a CM energy of $170\GeV$, the prediction in Figure~\ref{fig:wwcs2000} 
is continued by {\sc GENTLE} \cite{gentle}, which does not include 
the non-universal electroweak corrections. In its new version {\sc GENTLE} 
is tuned to reproduce the DPA prediction of
{\sc RacoonWW} and {\sc YFSWW3} on the total cross section at LEP2
within a few per mill 
(see \citere{lep2mcws}).

\section{Universal electroweak corrections --- improved Born
approximation}

\subsection{Preliminaries}

Universal radiative corrections are those parts of the full correction
that are connected to specific subprocesses, such as collinear photon
emission or running couplings, and lead to characteristic enhancement
factors. Owing to their universality such corrections are often related to
the lowest-order matrix element of the underlying process. In the
following we construct an {\it improved Born approximation} (IBA) for
the processes $\eeWWffff$ that is based on universal corrections only.
For the production subprocess the IBA closely follows the
approximation formulated in \citere{bo92} for on-shell W-pair
production. For the W~decay the IBA is identical with the
lowest-order prediction in the $\GF$ scheme, as suggested in  
\citere{rcwdecay1}.

In order to define the IBA, we first need the lowest-order matrix element
of the process
\beqar\label{eq:ee4f}
\Pep(p_+,\si_+)+\Pem(p_-,\si_-) &\;\to\;&
\PWp(k_+,\la_+)+\PWm(k_-,\la_-)
\nn\\
&\;\to\;&
f_1(k_1,\si_1)+\bar f_2(k_2,\si_2)+f_3(k_3,\si_3)+\bar f_4(k_4,\si_4).
\hspace{2em}
\eeqar
The arguments label the momenta $p_\pm$, $k_i$ and helicities
$\si_i=\pm1/2$, $\la_j=0,\pm1$ of the corresponding particles. 
The cross section that is defined by including only the
so-called {\it signal diagrams} for W-pair-mediated four-fermion
production, which are shown in \reffi{fig:sigdiags}, is called CC03
cross section%
\footnote{Of course, the CC03 cross section is a 
non gauge-invariant
quantity. However, evaluated in the `t~Hooft--Feynman gauge it
approximates the full cross sections for W-pair-mediated $4f$ production
very well, as long as no electrons or positrons 
are in the final state. Therefore,
the CC03 cross section is widely used in the literature (see also
\citeres{lep2repWcs,lep2mcws}).}. 
\bfi
\centerline{
\begin{picture}(135,80)(0,0)
\ArrowLine(30,40)(10,70)
\ArrowLine(10,10)(30,40)
\Vertex(30,40){1.2}
\Photon(30,40)(70,40){2}{4}
\Vertex(70,40){1.2}
\Photon(70,40)(100,60){2}{4}
\Photon(70,40)(100,20){2}{4}
\Vertex(100,60){1.2}
\Vertex(100,20){1.2}
\ArrowLine(100,60)(120,70)
\ArrowLine(120,50)(100,60)
\ArrowLine(100,20)(120,30)
\ArrowLine(120,10)(100,20)
\put(-5,58){$\Pep$}
\put(-5,12){$\Pem$}
\put(40,27){$\gamma,Z$}
\put(75,55){$W$}
\put(75,15){$W$}
\put(125,70){$f_1$}
\put(125,45){$\bar f_2$}
\put(125,30){$f_3$}
\put(125, 5){$\bar f_4$}
\end{picture}
\hspace{3em}
\begin{picture}(105,80)(0,0)
\ArrowLine(40,60)(10,70)
\ArrowLine(40,20)(40,60)
\ArrowLine(10,10)(40,20)
\Vertex(40,60){1.2}
\Vertex(40,20){1.2}
\Photon(40,60)(70,60){2}{3.5}
\Photon(40,20)(70,20){2}{3.5}
\Vertex(70,60){1.2}
\Vertex(70,20){1.2}
\ArrowLine( 70,60)( 90,70)
\ArrowLine( 90,50)( 70,60)
\ArrowLine( 70,20)( 90,30)
\ArrowLine( 90,10)( 70,20)
\put(-5,58){$\Pep$}
\put(-5,12){$\Pem$}
\put(45,35){$\nu_{\Pe}$}
\put(50,68){$W$}
\put(50, 5){$W$}
\put( 95,70){$f_1$}
\put( 95,45){$\bar f_2$}
\put( 95,30){$f_3$}
\put( 95, 5){$\bar f_4$}
\end{picture} }
\caption{Lowest-order signal diagrams for $\eeWWffff$}
\label{fig:sigdiags}
\efi
Note that the masses of the external fermions (not the ones in closed 
fermion loops) are neglected whenever possible.
In the absence of photon radiation this, in particular, implies that
we have helicity conservation for 
the initial $\Pep\Pem$ system, i.e.\ only the combination $\sigma_-=-\sigma_+$ 
contributes, and we can define $\sigma=\sigma_-=-\sigma_+$.
For definite electron helicity $\sigma$,
the lowest-order CC03 matrix element is given by
\beq
\label{eq:MbornCC03}
\M^{\eeWWffff,\si}_{\Born,\CCOT}(p_+,p_-,k_+,k_-,k_+^2,k_-^2)=
\sum_{n=1}^{3} F^\sigma_{n,\Born}(s,t) 
\M^\sigma_n(p_+,p_-,k_+,k_-,k_+^2,k_-^2),
\eeq
where $\M^\sigma_n$ are so-called standard matrix elements (SME) containing
the spinor chains of the external fermions, and
$F^\sigma_{n,\Born}(s,t)$ are invariant functions containing couplings
and propagator factors. In lowest order only three SME and invariant
functions contribute.
Following the notation and conventions of \citere{de00}, these read
($\omega_\pm=(1\pm\ga_5)/2$)
\beqar
\M^\sigma_1 &=& \bar v(p_+)\dsl\veps^*_+(\dsl{k}_+ -\dsl p_+)
\dsl\veps^*_- \omega_\sigma u(p_-),
\nn\\
\M^\sigma_2 &=& \bar v(p_+)\text\frac{1}{2}(\dsl{k}_+ -\dsl{k}_-)
\omega_\sigma u(p_-) (\veps^*_+\veps^*_-),
\nn\\
\M^\sigma_3 &=& \bar v(p_+)\dsl\veps^*_+\omega_\sigma u(p_-)
(\veps^*_- k_+)
-\bar v(p_+)\dsl\veps^*_-\omega_\sigma u(p_-)
(\veps^*_+ k_-)
\label{eq:smes}
\eeqar
with ``effective W-polarization vectors''
\beqar
\veps^{*,\mu}_+ &=& \frac{e}{\sqrt{2}\sw} \,
\frac{1}{k_+^2-\MW^2+\ri\MW\GW} \,
\bar u(k_1)\gamma^\mu\omega_- v(k_2),
\nn\\
\veps^{*,\mu}_- &=& \frac{e}{\sqrt{2}\sw} \,
\frac{1}{k_-^2-\MW^2+\ri\MW\GW} \,
\bar u(k_3)\gamma^\mu\omega_- v(k_4),
\label{eq:effpols}
\eeqar
and
\beqar\label{eq:Fborn}
F^\sigma_{1,\Born}(s,t) &=& \frac{e^2}{2\sw^2 t} \delta_{\sigma-}, 
\nn\\
F^\sigma_{3,\Born}(s,t) &=& -F^\sigma_{2,\Born}(s,t) = \frac{2e^2}{s}
-\frac{2e^2}{s-\MZ^2}\biggl(1-\frac{\delta_{\sigma-}}{2\sw^2}\biggr).
\eeqar
The actual values of the input parameters $e$, $\MW$, $\MZ$, and $\sw$
depend on the input-parameter scheme. In the $\GF$-scheme the
electromagnetic coupling $e$ is deduced from the Fermi constant $\GF$
using the tree-level relation $e^2=4\sqrt{2}\GF\MW^2\sw^2$,
and the weak mixing angle is fixed by the gauge-boson masses, which are
independent input parameters, $\sw^2=1-\MW^2/\MZ^2$.

Before we define the IBA we comment on the calculation of the full
factorizable one-loop correction in DPA%
\footnote{
In DPA the virtual one-loop correction consists of {\it factorizable}
and {\it non-factorizable} contributions. 
The factorizable corrections are the ones
that are related to the W-pair-production and W-decay subprocesses.
The non-factorizable corrections 
comprise the remaining doubly-resonant virtual corrections and include
all diagrams with photon exchange between the production and decay 
subprocesses.}, 
which is described in 
\citere{de00}, and its relation to the decomposition \refeq{eq:MbornCC03}.
In this case six independent SME contribute for each 
value of $\sigma$, and the functions $F^\sigma_n$ contain standard loop
integrals. Moreover, in order to guarantee the gauge invariance of the
corrections, which is mandatory for consistency, it is necessary to
perform an on-shell projection of the external fermion momenta $k_i$.
This means that the $k_i$ are changed to related momenta $\kon_i$ in
such a way that $\kon_\pm^2=\MW^2$. The off-shell values $k_\pm^2$ are
kept only in the propagator factors of \refeq{eq:effpols}.%
\footnote{This on-shell projection also renders the CC03 cross section
gauge-invariant, leading to the so-called DPA Born cross section.
However, the DPA Born cross section is a much worse approximation for
the $4f$ cross section than the CC03 variant 
(see also \citeres{lep2mcws,Be98}).}

\subsection{Improved Born approximation}

The first step in the construction of the IBA consists in a modification
of the Born matrix element in such a way that the universal
renormalization effects induced by the running of $\alpha$ and by
$\Delta\rho$ are absorbed. This is achieved \cite{bo92} by the replacements
\beq
\frac{e^2}{\sw^2} \;\to\; 4\sqrt{2}\GF\MW^2, \qquad
e^2 \;\to\; 4\pi\alpha(s)
\eeq
in the lowest-order functions $F^\sigma_{i,\Born}$ of \refeq{eq:Fborn},
which implies that weak-isospin exchange involves the coupling
$\GF\MW^2$ and pure photon exchange the coupling $\alpha(s)$.
The running of the electromagnetic coupling is induced by light
(massless) charged fermions only, i.e.\ we evaluate $\alpha(s)$ by
\beq
\disp
\alpha(s) = \frac{\alpha(\MZ^2)}
        {1-\frac{\alpha(\MZ^2)}{3\pi}\ln(s/\MZ^2) 
        \sum_{f\ne\Pt}N^{\mathrm{c}}_f Q_f^2}
\eeq
with the value $\alpha(\MZ^2)=1/128.887$ taken from the fit \cite{alphamz}
of the hadronic vacuum polarization to the empirical ratio
$R=\sigma(\Pep\Pem\to\mbox{hadrons})/\sigma(\Pep\Pem\to\mu^+\mu^-)$.
Thus, the basic matrix element for the IBA reads
\beq
\label{eq:MIBA}
\M^{\eeWWffff,\si}_{\IBA}(p_+,p_-,k_+,k_-,k_+^2,k_-^2)=
\sum_{n=1}^{3} F^\sigma_{n,\IBA}(s,t) 
\M^\sigma_n(p_+,p_-,k_+,k_-,k_+^2,k_-^2)
\eeq
with
\beqar\label{eq:FIBA}
F^\sigma_{1,\IBA}(s,t) &=& \frac{2\sqrt{2}\GF\MW^2}{t} \delta_{\sigma-}, 
\nn\\
F^\sigma_{3,\IBA}(s,t) &=& -F^\sigma_{2,\IBA}(s,t) =
\frac{4\sqrt{2}\GF\MW^2}{s-\hat\MZ^2} \delta_{\sigma-}
-\frac{8\pi\alpha(s)\hat\MZ^2}{s(s-\hat\MZ^2)}.
\eeqar
Note that we have used the complex Z-boson mass
$\hat\MZ^2=\MZ^2-\ri\MW\GZ$ in order to regularize the Z~resonance
below the W-pair-production threshold; otherwise the ISR convolution
over the reduced CM energy would lead to complications (see below).

Another important virtual correction is induced by the Coulomb
singularity near the W-pair-production
threshold. We include this effect in the
calculation of the ``hard'' IBA cross section 
$\hat\sigma_\IBA^{\eeWWffff}$,
\beq
\int \rd\hat\sigma_\IBA^{\eeWWffff}(p_+,p_-) =
\frac{1}{2s} \int\rd\Phi_{4f}\, |\M_{\IBA}^{\eeWWffff}|^2 \,
\left[1+\delta_{\Coul}(s,k_+^2,k_-^2) \, g(\bar\beta)\right],
\eeq
where the correction factor $\delta_{\Coul}$ is given by
\cite{coul,nfc1a}
\beqar
\delta_{\Coul}(s,k_+^2,k_-^2) &=& \frac{\alpha(0)}{\bar\beta}
\Im\left\{\ln\left(\frac{\beta-\bar\beta+\Delta_M}
        {\beta+\bar\beta+\Delta_M}\right)\right\},\nl
\bar\beta &=& \frac{\sqrt{s^2+k_+^4+k_-^4-2sk_+^2-2sk_-^2-2k_+^2k_-^2}}{s},\nl
\beta &=&  \sqrt{1-\frac{4(\MW^2-\ri\MW\GW)}{s}}, \qquad
\Delta_M = \frac{|k_+^2-k_-^2|}{s}
\eeqar
with the fine-structure constant $\alpha(0)$.
The auxiliary function 
\beq
g(\bar\beta) = \left(1-\bar\beta^2\right)^2
\eeq
restricts the impact of $\delta_{\Coul}$ to the threshold region where 
it is valid. Its actual form (and its occurrence) is somewhat ad hoc but
justified by a numerical comparison to the full ${\cal O}(\alpha)$ correction.
Omitting this factor would lead to a constant positive correction 
of a few per mill above threshold, although 
the correct non-universal correction is even negative.

The last ingredient in the IBA is the leading-logarithmic contribution
induced by initial-state radiation (ISR). We follow the
structure-function approach \cite{sf}, where the full IBA cross section
$\sigma_{\IBA}$ reads
\newcommand{\LL}{\mathrm{LL}}
\beq\label{sigmaIBA}
  \int \rd\sigma_{\IBA} =
  \int^1_0 \rd x_1 \int^1_0 \rd x_2 \,
  \Gamma_{\Pe\Pe}^{\LL}(x_1,Q^2)\Gamma_{\Pe\Pe}^{\LL}(x_2,Q^2)
  \int \rd\hat\sigma_\IBA^{\eeWWffff}(x_1 p_+,x_2 p_-).
\eeq
The structure functions $\Gamma_{\Pe\Pe}^{\LL}(x,Q^2)$ include the
leading logarithms $[\alpha\ln(Q^2/\Me^2)]^n$ up to order $n=3$, and the 
soft-photon effects are exponentiated; the explicit expressions can also
be found in \citeres{lep2repWcs,de00}. 
The QED splitting scale $Q^2$ is not fixed in
leading-logarithmic approximation and has to be set to a typical
momentum scale of the process. It can be used to adjust the IBA to the
full correction, but also to estimate the intrinsic uncertainty of the
IBA by choosing different values for $Q^2$.

Finally, we have to fix the W-boson width $\GW$ in the evaluation of the IBA. 
In order to avoid any kind of mismatch with the decay, $\GW$ should be
calculated in lowest order using the $\GF$ scheme. This choice
guarantees that the ``effective branching ratios'', which result after
integrating out the decay parts, add up to one when summing over all
channels. Of course, if naive QCD corrections are taken into account by
multiplying with $(1+\alpha_{\mathrm{s}}/\pi)$ for each hadronically
decaying W~boson, these QCD factors also have to be included in the
calculation of the total W~width.

Note that unlike the full one-loop calculation in DPA, the IBA is also
applicable near the W-pair production threshold, since no pole
expansion is involved.

\section{\sloppy Comparison of the improved Born approximation with
state-of-the-art results}

\subsection{Total cross section}

In order to investigate the reliability of the IBA defined in
\refeq{sigmaIBA}, we have implemented this IBA in the Monte
Carlo program {\sc RacoonWW}, which provides state-of-the-art
predictions for the full ${\cal O}(\alpha)$ corrections in DPA, as
discussed above. For the following numerical evaluations we have adopted
the input-parameter set of \citeres{lep2mcws,de00}.

\begin{figure}
\setlength{\unitlength}{1cm}
\centerline{
\begin{picture}(10.5,8.0)
\put(-3.0,-19.0){\includegraphics{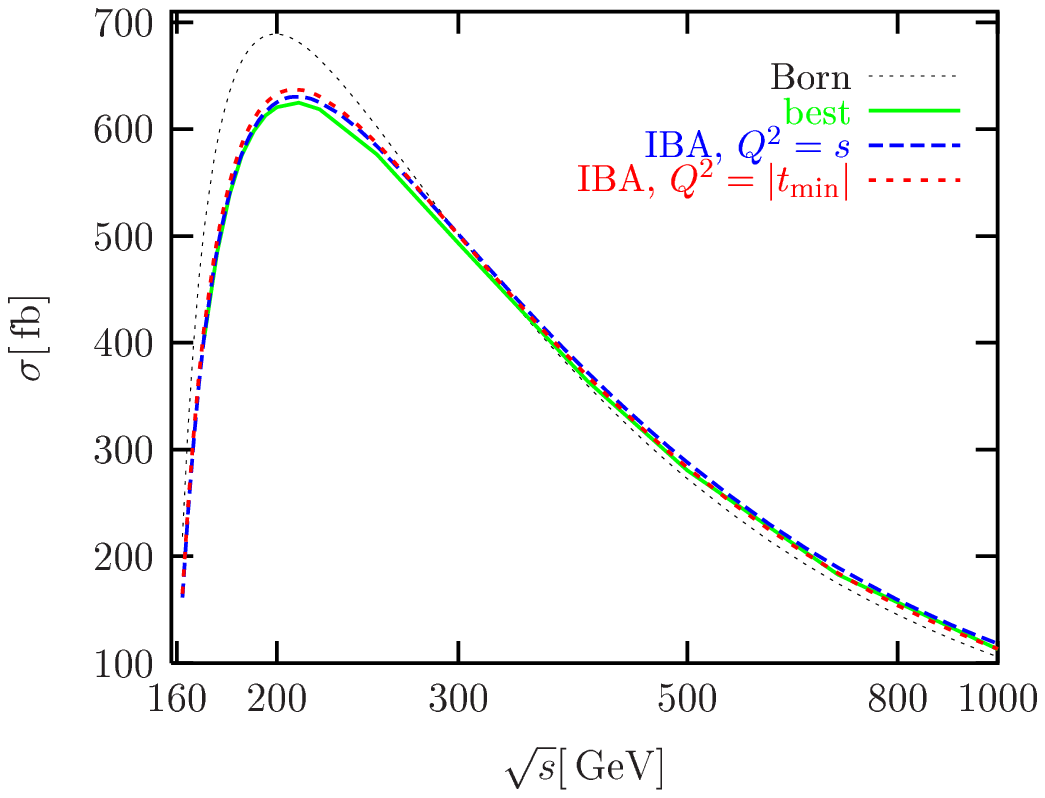}}
\end{picture}
}
\caption{Predictions for the total cross section for the process
$\Pep\Pem\to\Pu\bar\Pd\mu^-\bar\nu_\mu$ based on various approximations
for radiative corrections}
\label{fig:IBAabs}
\vspace*{2em}
\centerline{
\begin{picture}(10.5,8.0)
\put(-3.0,-19.0){\includegraphics{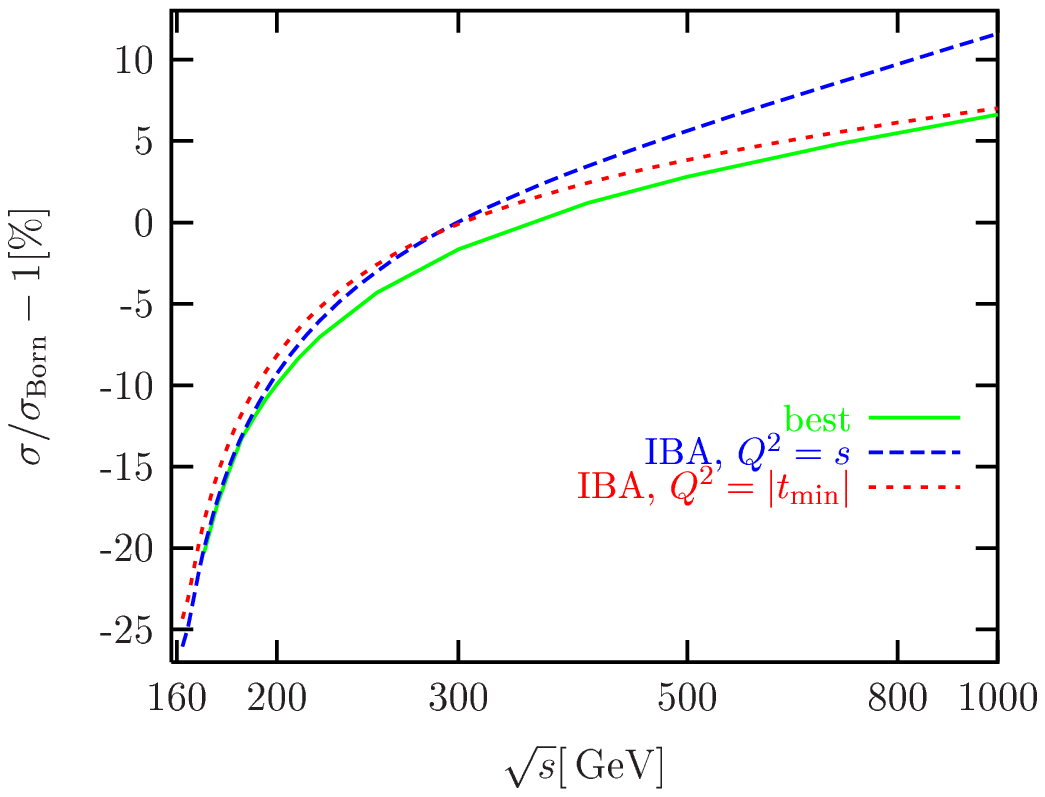}}
\end{picture}
}
\caption{Relative corrections to the total cross section for the process
$\Pep\Pem\to\Pu\bar\Pd\mu^-\bar\nu_\mu$ in various approximations}
\label{fig:IBArel}
\end{figure}

Figure~\ref{fig:IBAabs} compares different predictions for the total 
cross section 
(without any phase-space cuts) for the semileptonic process 
$\Pep\Pem\to\Pu\bar\Pd\mu^-\bar\nu_\mu(\gamma)$ for CM energies $E_{\CM}$
up to $1\TeV$.
The IBA is evaluated for the two different scales 
$Q^2=s=E_{\CM}^2$ and 
$Q^2=|t_{\min}|=E_{\CM}^2\left(1-\sqrt{1-4\MW^2/E_{\CM}^2}\right)/2-\MW^2$,
and ``best'' labels the 
{\sc RacoonWW} prediction including all 
universal and non-universal corrections as described in detail 
in \citere{de00}.
The motivation for $Q^2=s$ is obvious; $Q^2=|t_{\min}|$ is motivated by
the fact that $t_{\min}$ corresponds to the minimal
momentum transfer in the $t$-channel diagrams for forward scattering
of on-shell W~bosons, which dominates the cross 
section. The comparison of the corresponding relative corrections 
(normalized to the CC03 Born cross section in $\GF$ scheme) is shown in 
\reffi{fig:IBArel}.

For LEP2 energies, \ie energies below $210\GeV$,
the difference between the two IBA versions reflects
the typical uncertainty of 1--2\% inherent in all predictions that
neglect non-universal electroweak corrections. It turns out that the IBA
with $Q^2=s$ is closer to the ``best'' prediction, with a maximal
deviation at the upper LEP2 energies: $\sim 0.6\%$ at $200\GeV$ and
$\sim 0.8\%$ at $210\GeV$. Note that the ``best''
prediction is not included below $170\GeV$, since the uncertainty of
all predictions based on a DPA formally runs out of control near the
W-pair-production
threshold. On the other hand, the IBA does not suffer from this
constraint. Since the IBA with $Q^2=s$ 
agrees with the ``best'' prediction near $170\GeV$ at the per-mill level
this IBA version is an appropriate
extrapolation of the ``best'' {\sc RacoonWW} prediction down to the
W-pair-production threshold. Of course, the theoretical uncertainty
below $170\GeV$ is then of the order of one to a few per cent.

For LC energies the IBA becomes more and more uncertain;
for $1\TeV$ the two IBA versions differ already by $\sim 5\%$. This
signals that non-leading electroweak corrections become more and more 
important. The dominant effects are due to Sudakov logarithms \cite{be93}
of the type $\alpha\ln^2(s/\MW^2)$ which originate
from the exchange of soft 
and collinear massive gauge bosons, i.e.\ W and Z~bosons. The IBA does
not account for these effects. Nevertheless the IBA with $Q^2=|t_{\min}|$
follows the ``best'' prediction within $\sim 1$--2\% even for high
energies. This is plausible, because the total cross section is
strongly dominated by the $t$-channel pole for forward scattering for
high energies,
and this contribution is well approximated by the IBA. 
Note, however, that the good agreement could not be
predicted without a comparison with the full DPA correction including
non-universal electroweak corrections. On the other hand, it can be
expected that the quality of the IBA with $Q^2=|t_{\min}|$ also
becomes worse if forward scattering is excluded or suppressed by 
phase-space cuts; this issue is further discussed below in the context
of differential distributions.

\subsection{Differential distributions}

In order to define differential distributions, the kinematic information
on the fermion momenta in 
$\eeWWffff(\gamma)$ is required. In the presence of
photon radiation, a consistent treatment of photons that are soft or
collinear to charged fermions is crucial. If such photons are not
recombined with the nearest charged fermion, i.e.\ if these
photon--fermion systems are not treated as single ``quasi-particles'', 
the bare fermion momenta in general lead to distributions that are not
IR-safe, i.e.\ they involve mass-singular logarithms of the form
$\alpha\ln m_f$. For fermions other than muons such effects are
definitely unphysical. In order to avoid such artifacts, we recombine
soft and collinear photons according to the procedure%
\footnote{In this approach, first
photons close to the beams are dropped in events, i.e.\ their momenta 
are set to zero. 
If the photon survives the cut to the beam, it is recombined with the 
charged fermion $f$ if $M_{f\gamma}<M_\recomb$,
where $f$ is the fermion with the smallest invariant mass
$M_{f\gamma}^2=(p_f+k_\gamma)^2$ with the photon. Finally, events are
discarded in which charged fermions are close to the beam. The size
of the recombination cut $M_\recomb$, thus, determines how many photons
are recombined with the charged fermions. In \citeres{lep2mcws,de00} the
two values $M_\recomb=5\GeV$ and $25\GeV$ are chosen, defining a
``bare'' and a ``calo(rimetric)'' setup, respectively.}
described in \citeres{lep2mcws,de00}.

\begin{figure}
\setlength{\unitlength}{1cm}
\centerline{
\begin{picture}(16,8.0)
\put(-5.3,-17.8){\includegraphics{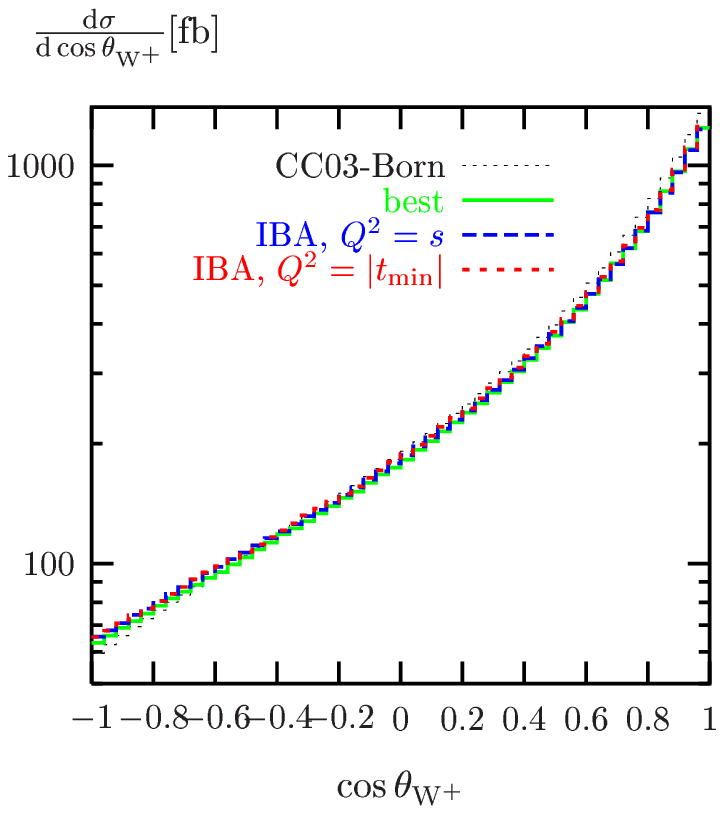}}
\put( 3.0,-17.8){\includegraphics{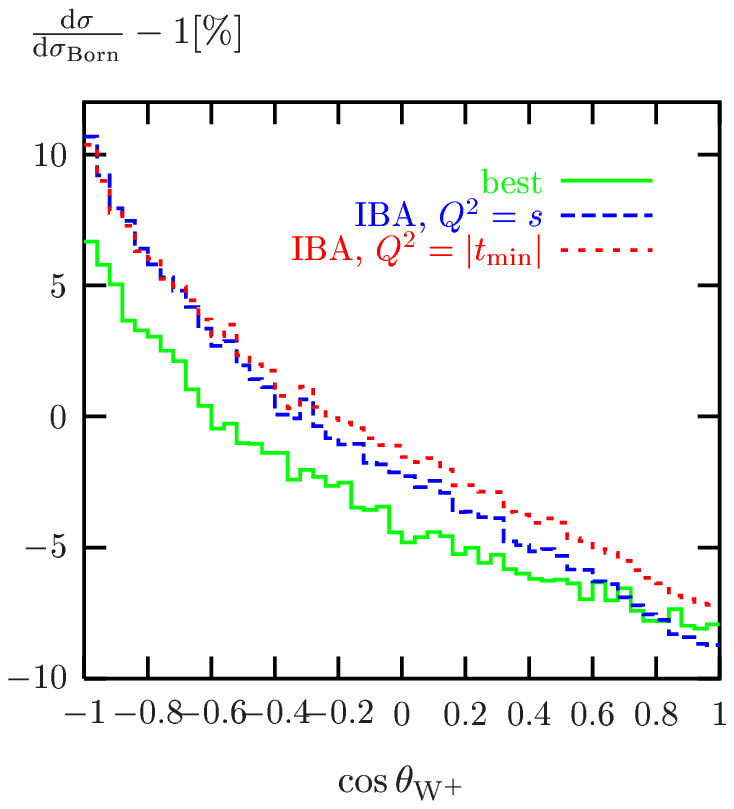}}
\end{picture}
}
\caption{Predictions for the $\PWp$-production-angle distribution (left) and 
corresponding relative corrections (right) for the process
$\Pep\Pem\to\Pu\bar\Pd\mu^-\bar\nu_\mu$ at $E_{\CM}=200\GeV$ 
based on various approximations for radiative corrections}
\label{fig:IBAct200}
\vspace*{4em}
\setlength{\unitlength}{1cm}
\centerline{
\begin{picture}(16,8.0)
\put(-5.3,-17.8){\includegraphics{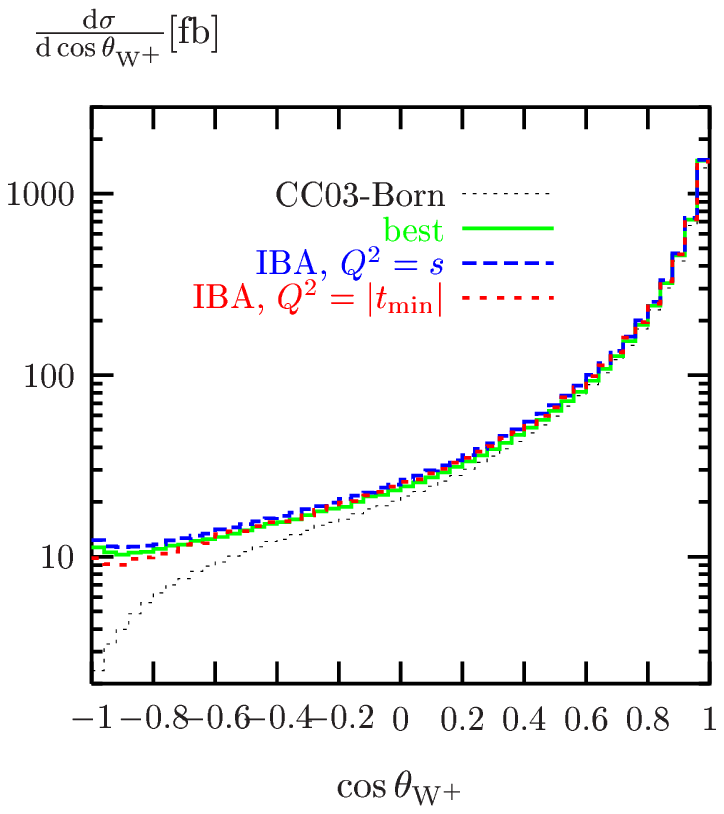}}
\put( 3.0,-17.8){\includegraphics{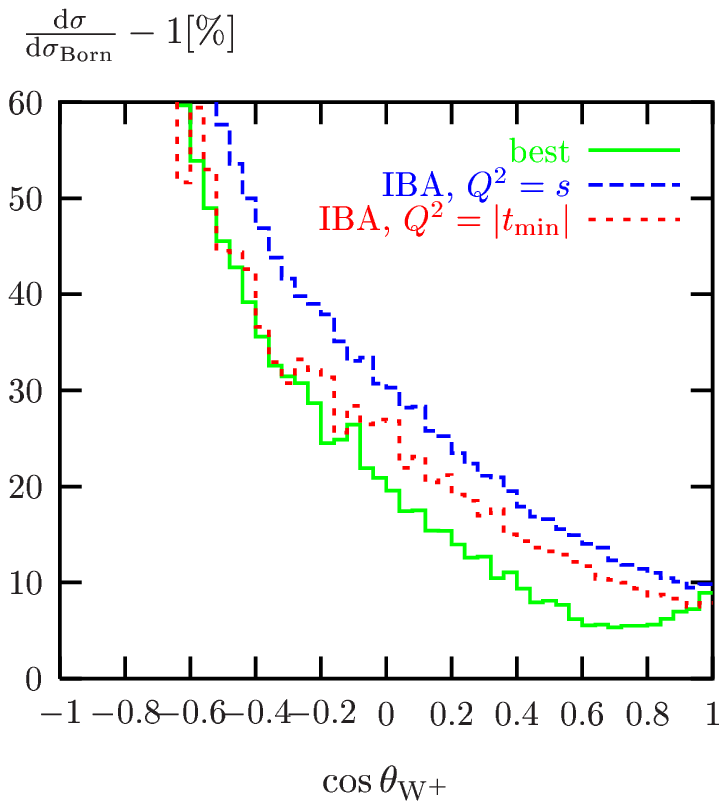}}
\end{picture}
} 
\caption{Predictions for the $\PWp$-production-angle distribution (left) and 
corresponding relative corrections (right) for the process
$\Pep\Pem\to\Pu\bar\Pd\mu^-\bar\nu_\mu$ at $E_{\CM}=500\GeV$ 
based on various approximations for radiative corrections}
\label{fig:IBAct500}
\end{figure}
In \reffis{fig:IBAct200} and \ref{fig:IBAct500} the full {\sc RacoonWW} 
and IBA 
predictions for the $\PWp$-pro\-duc\-tion-angle distribution are compared 
for the process $\Pep\Pem\to\Pu\bar\Pd\mu^-\bar\nu_\mu$ at the typical LEP2
energy of $E_{\CM}=200\GeV$ and the LC energy $E_{\CM}=500\GeV$.
The uncertainty of the IBA predictions induced by the QED splitting
scale $Q$ is about 1--2\% and $\sim 5\%$ for LEP and LC energies, respectively.
The deviation of the IBA prediction from the full result is 
up to $\sim 5\%$ and $\sim 5$--10\%,
where the agreement is best for forward scattering
($\cos\theta_{\PWp}\to 1$), as anticipated above. The IBA uncertainty
and the deviation from the full result further grow with increasing
energy. This comparison is performed using the ``calo'' setup for the
photon recombination of photons, but the sensitivity of the
W-production-angle distribution to the recombination procedure is very
weak (see \citeres{lep2mcws,de00}).

\begin{figure}
\setlength{\unitlength}{1cm}
\centerline{
\begin{picture}(16,8.0)
\put(-5.3,-17.8){\includegraphics{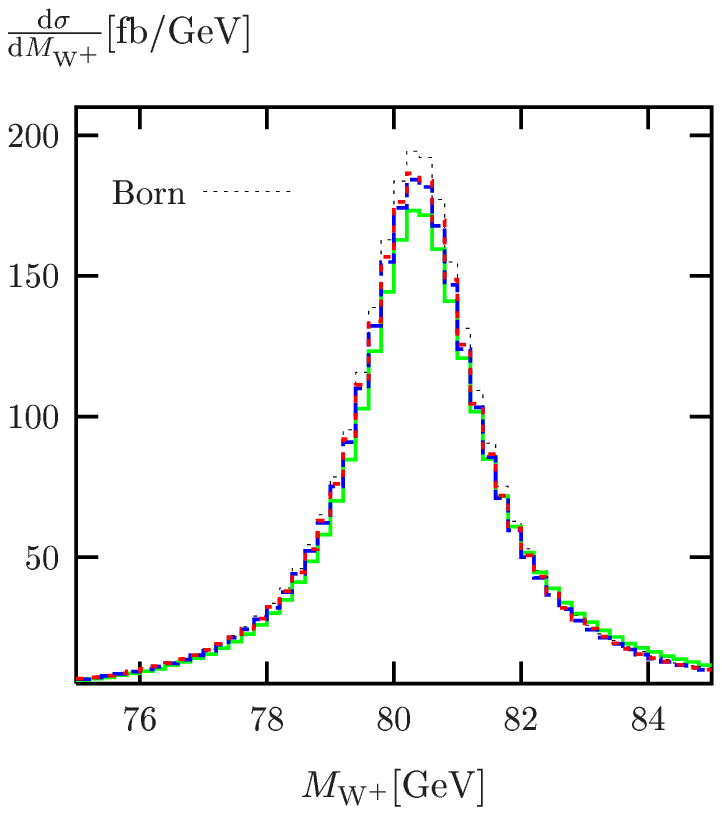}}
\put( 3.0,-17.8){\includegraphics{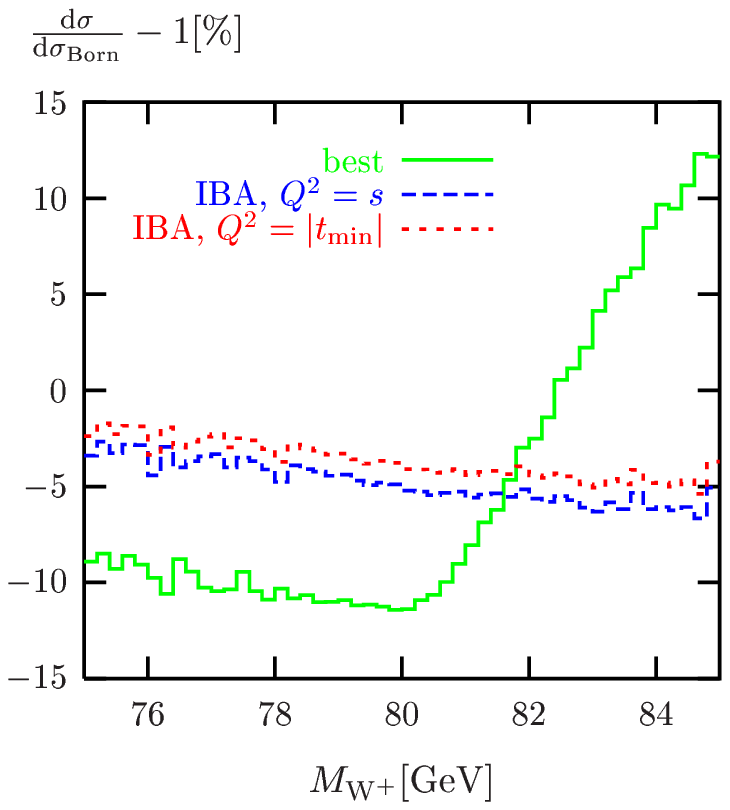}}
\end{picture}
} 
\caption{Predictions for the $\PWp$-invariant-mass distribution (left) and 
corresponding relative corrections (right) for the process
$\Pep\Pem\to\Pu\bar\Pd\mu^-\bar\nu_\mu$ at $E_{\CM}=200\GeV$ 
based on various approximations for radiative corrections, using the
``calo'' setup for photon recombination}
\label{fig:IBAmwpcalo}
\vspace*{4em}
\setlength{\unitlength}{1cm}
\centerline{
\begin{picture}(16,8.0)
\put(-5.3,-17.8){\includegraphics{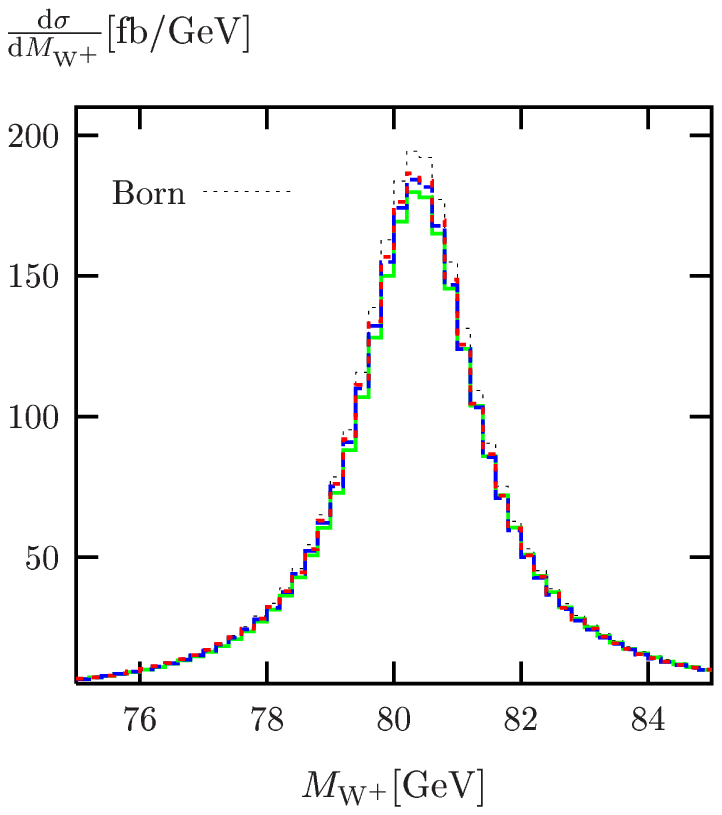}}
\put( 3.0,-17.8){\includegraphics{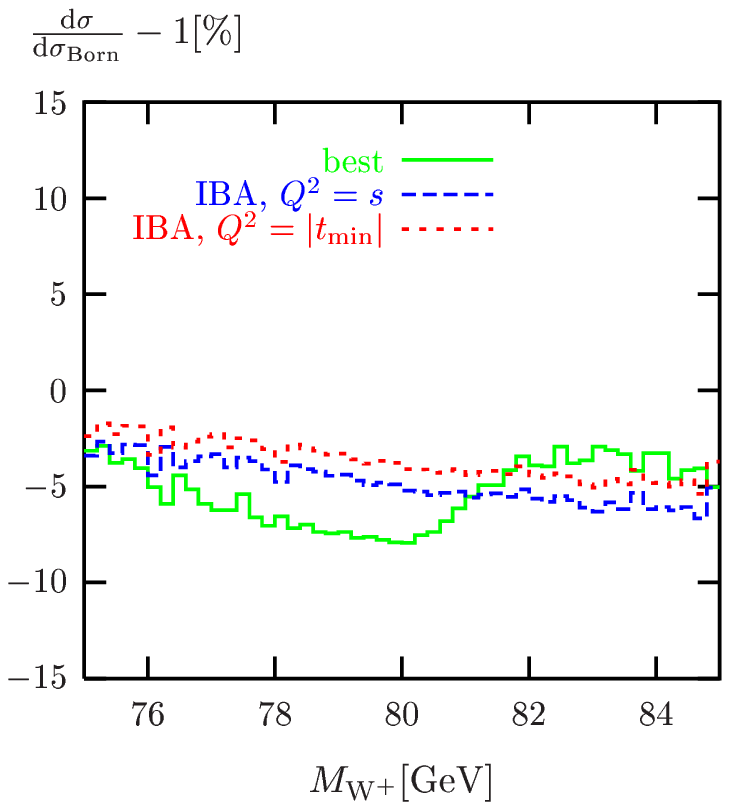}}
\end{picture}
} 
\caption{Predictions for the $\PWp$-invariant-mass distribution (left) and 
corresponding relative corrections (right) for the process
$\Pep\Pem\to\Pu\bar\Pd\mu^-\bar\nu_\mu$ at $E_{\CM}=200\GeV$ 
based on various approximations for radiative corrections, using the
``bare'' setup for photon recombination}
\label{fig:IBAmwpbare}
\end{figure}

The sensitivity to photon recombination is maximal in the invariant-mass
distributions of the reconstructed W~bosons, which can be seen by
comparing \reffis{fig:IBAmwpcalo} and \ref{fig:IBAmwpbare}.
The full correction shows a very strong dependence on the recombination
procedure, which was discussed in \citere{de99a} in detail.
The more inclusive recombination (``calo'') leads to large
positive corrections above resonance and thus to a shift of the
resonance to the right, which can be of the order of some $10\MeV$ 
\cite{de99a}.
Since this distortion of the W~line shape is
mainly induced by final-state radiation and radiation off the W~bosons,
the IBA, as defined above, does not account for this effect. It is
obvious that the W-invariant-mass distributions can only be properly
described if photon radiation from the W-decay processes is taken into
account 
properly. Figures \ref{fig:IBAmwpcalo} and \ref{fig:IBAmwpbare}
refer to the LEP energy $E_{\CM}=200\GeV$, but this conclusion is, of
course, valid for all energies.

\section{Conclusions}

Electroweak radiative corrections to $\eeWWffff$ typically amount to
${\cal O}(10\%)$ at LEP2 energies and further increase for higher
energies. We have explicitly given analytical results for the universal
process-independent corrections, which include effective
coupling constants, the Coulomb singularity near the W-pair-production
threshold,
and leading ISR effects. They have been implemented in the Monte Carlo
generator {\sc RacoonWW}, which calculates the full ${\cal O}(\alpha)$
corrections in double-pole approximation. Using this program a 
comparison between universal effects and the full correction has been
presented.

For LEP2 energies the universal corrections are dominant, and the
remaining non-universal contributions reduce the total W-pair cross
section by 1--2\%. In angular distributions non-universal effects can
reach several per cent, mainly in regions where the cross section is
small. The radiative corrections to W-invariant-mass distributions
lead to a distortion of the W~resonance, which is mainly due to photon
radiation off the charged final-state fermions and off the W~bosons.
This line-shape distortion is not accounted for by
the above-mentioned universal effects.

For LC energies, i.e.\ energies up to the TeV range, non-universal
effects become more and more important. While the universal effects
still describe W-pair production in the forward region within some per
cent, non-universal corrections reach the order of several 10\% for
intermediate and large W-production angles. 

\Acknowledgments We would like to thank the organizers of the
conference, in particular Howard Haber, for providing a very pleasant
atmosphere during the whole conference. Two of us (A.D.\ and S.D.)
would like to thank the organizers for financial support. This work
was supported in part by the Swiss Bundesamt f\"ur Bildung und
Wissenschaft, by the European Union under contract
HPRN-CT-2000-00149,
by the U.S. Department of Energy
under grant DE-FG02-91ER40685 and by the U.S. National Science Foundation
under grant PHY-9600155.


\end{document}